# Integration of the Equation of the artificial Earth's Satellites Motion with Selection of Runge-Kutta-Fehlberg Schemes of Optimum Precision Order

Atanas Marinov Atanassov

Solar Terrestrial Influences Institute, Bulgarian Academy of Sciences,
Stara Zagora Department, P.O. Box 73, 6000 Stara Zagora, Bulgaria
E-mail: At_M_Atanassov@yahoo.com

**Abstract.** *An approach is treated for numerical integration of ordinary differential equations systems of the first order with choice of a computation scheme, ensuring the required local precision. The treatment is made on the basis of schemes of Runge-Kutta-Fehlberg type. Criteria are proposed as well as a method for the realization of the choice of an "optimum" scheme. The effectiveness of the presented approach to problems in the field of satellite dynamics is illustrated by results from a numerical experiment. These results refer to a case when a satisfactory global stability of the solution for all treated cases is available. The effectiveness has been evaluated as good, especially when solving multi-variable problems in the sphere of simulation modelling.*

**Introduction.** The practice of computing often demands the solution of ordinary differential equation systems (ODES) of the first order:

(1)     $\dfrac{dy_i}{dx} = \varphi_i(x, y_i)$

with initial conditions $y_i(x_0) = y_i^0$. $y_i(x_0) = y_i^0$. The type of the right side of (1) determines the problem as stiff or non-stiff.

Different groups of methods exist for numerical solution of (1) in cases of non-stiff problems. First, there are ones that are with general character, applicable to a wide class of problems and there are others, which are specialized, oriented to specific problems. The latter have certain advantages, yet they require development for each specific case. An example of the second type of methods is the so-called recurrent power series [1]. The methods with more common application are divided mainly into one-step and multiple-step ones. [2].

As far as the order of the equations systems of higher order is subject to lowering, the methods for first-order systems are always applicable. There are also methods, which are applied without preliminary order lowering. As a rule they are more efficient in cases when they can be applied [3, 4] but Nystrom's schemes [5] (which are the most popular) have a more restricted area of application since they require the right part to be independent from the first derivative of the dependent variable *y*.

In order to minimize the error and the computational expenses for each integration step, used to make the computations, major importance is attributed to the step size for numerical integration and the order of the computational schemes. Approaches exist for determination of optimum step size, which are possible both with the one-step and the multi-step methods. The change of the step in the one-step methods is easier while the multi-step methods require re-computation of the function derivative values in new points and produce complications [6]. The possibility to change the order (line) of the integration method in some multi-step methods [2, 7] attracts attention.

Two basic final approaches exist in the numerical integration of ODES. The first one is connected with integration through optimal step selection. The optimal step size is determined on the basis of error assessment; in general, at smaller error the step increases and vice versa. Optimization of the computational expenses is achieved along with ensuring global stability of the numerical solution at the end of the integration interval.

The second approach requires finding the solution in equidistant values by the independent variable. The integration by a constant step, however, doesn't always meet the requirement for sufficient local error, connected with the type of the functions on the right side of the equations, hence it can influence the solution stability.

There exists, however, a possibility during integration with optimal step to obtain the solution in desired points on the independent variable on the basis of interpolation. In addition – special methods exist, combining the numerical integration and interpolation, with which the solution is obtained in arbitrary points in a natural way with increased computational efficiency [8, 9].

Different methods and programs-integrators of common differential equations are developed and their efficiency has been examined. As a result of their large number we'll mention only a few, having in common with the evident one-step methods of the Runge-Kutta type [10, 11, 12, 13]. Although the efficiency estimations show some advantages in behalf of one or another method and computer programs, when solving test problems there isn't any certainty as for which method is the most suitable one.

A number of methodological groups exist for numerical solution of (1). The one-step methods of Runge-Kutta type [1] are characterised by adaptability and easy programme realization. Different computing schemes, corresponding to those methods are known. The advantages of the schemes of the Runge-Kutta-Fehlberg (RKF) type [2,3] are due to the fact that with minimum additional computations, two solutions are simultaneously obtained with different precision:

$$(2) \quad y_i = y_i^0 + h\sum_{k=0}^{n} c_k g_k^i + O(h^{p+1}), \quad \overline{y_i} = y_i^0 + h\sum_{k=0}^{n+1} \overline{c_k} g_k^i + O(h^{p+2})$$

where $\quad g_0^i = \varphi_i^0(x_0, y_i^0), \quad g_k^i = \varphi_i\left(x_0 + a_k.h, y_i^0 + h\sum_{k=0}^{k-1} b_{kj} g_j\right).$

The difference between the two solutions $\overline{y}_i - y_i$ gives the exact value of the main member of the local error $\overline{O}(h^{p+1})$ for a scheme of the p$^{-th}$ order, on the basis of which the local error can be estimated. The classical RKF schemes were followed by later schemes with enhanced efficiency as well as by a higher order of the solution precision [16, 17, 18, 19]. In cases of computations made on the basis of a scheme of the p$^{-th}$ order, by changing the integration step in definite limits, the necessary local precision is obtained and hence - a certain stability of the solution.

**Formulation of the Problem.** In the integration region along the independent variable x, the local error is a variable quantity and depends on step h, on the p$^{-th}$ order of the integration scheme and on the type of the functions on the right side of (1). The minimization of the local error by means of stepsize control is not the only possible. Practically, the minimization of the local error by a stepsize control is not always suitable. Instead, we can use an integration scheme of the lowest possible order, which provides the necessary local precision in integration with constant step. In this way the computations can turn out to be considerably less than if a scheme is used of an order, providing precision for the entire integration region along an independent variable. The issue for selection of optimal order of the integration schemes is dated far back [20, 21]. This is possible even more in case when parallel integration of several ODES is necessary when the solutions have different character and are obtained with different local precision.

In the classical one-step methods of the Runge-Kutta type, the choice of a scheme with sufficient local precision is a problem, connected with the possibility for evaluation of the local error. In this aspect, the methods of the Runge-Kutta-Fehlberg type have advantages, which allow choosing a strategy for selection of the optimum integration scheme [22]. RKF schemes are known of 1/2, 2/3, 3/4, 4/5, 5/6 and 7/8 order [14, 15]. In the schemes of 1/2 up to 4/5 order, the functions on the right side of (1) are computed in 3 up to 6 intermediate points, respectively, and for the schemes of 5/6 and 7/8 order - in 8 and 13, respectively. Special methods are known for integration of ODES with variable order of precision [2, 7] but instead we shall examine the possibility to use the RKF schemes.

Let's consider a possibility to investigate the effectiveness in the selection of an optimum scheme, connected with the integration of the equation of the artificial Earth's satellites motion (AES). The most common form of this equation is [23]:

$$(3) \quad m\frac{d^2\vec{r}}{dt^2} = -G\frac{mM}{r^3}\vec{r} + \vec{f}$$

In (3) $\vec{r}$ is the satellite radius vector, m- its mass, M- the Earth mass, G- the gravitation constant, $\vec{f}$ - the disturbing forces, which determine the motion model and t- the time. The vector equation (3) is usually solved by decreasing the order to a system of two equations:

$$(4) \quad m\frac{d\vec{V}}{dt} = -G\frac{mM}{r^3}\vec{r} + \vec{f}$$

$$\frac{d\vec{r}}{dt} = \vec{V}$$

In (4) $\vec{V}$ is the velocity vector. The conditions, which specify the numerical integration local error of (4) at a constant step $\Delta t$ and a scheme of the $p^{-th}$ order are different for the separate sections of the trajectory.

**A Strategy for Choosing the Integration Scheme.** The choice of an integration scheme aims at providing not minimum at any case, but sufficient local error with which the numerical solution will be steady within a given interval, with minimum computation expenses.

In [22, 24] we examined the following strategy of choosing an optimum integration scheme for numerical integration based on the following cases:

a) For some functions $\varphi_i$ the local error $\overline{O^i}$ has a minimum value, bigger than the necessary one; a scheme of a higher order is now used and the computations are repeated.

b) smaller errors $\overline{O^i} < S^i \cdot O^i_{max}$ are obtained for all functions where $S^i$ are rough estimations of the relations $O^i(h^{p+2})/O^i(h^{p+1})$; In this case it is reasonable to pass to a scheme of a lower order.

Here it should be kept in mind that the estimation of $S^i$ is an essential problem. The experience shows that the above-stated contradicting requirements can be verified more precisely in the specific case of solving (4). This is made when instead of the local errors $\overline{O^i}$, i=1,..., 6, an evaluation of error $\overline{O} = \sqrt{\overline{O_x}^2 + \overline{O_y}^2 + \overline{O_z}^2}$ is made. Besides, when proceeding with an integration scheme of a higher order, the relative error value should be preserved. Later on, this value can be used as a lower limit, the passing of which should be accompanied by selection of a scheme of a lower order. This solves the problem with the difficulties in estimating $S^i$. The strategy of choosing a scheme can be presented, as follows:

a') $\overline{O} > \varepsilon \cdot \Delta r$, where $\Delta r = \sqrt{\Delta x^2 + \Delta y^2 + \Delta z^2}$, $\varepsilon$ is the value of the relative error selected with a view to reaching a global stability of the solution; then it is proceeded with the selection of a scheme of a higher order and when reaching a suitable scheme, the value of $\overline{O}$ is included in $\tilde{O}$.

b') $\overline{O} < \tilde{O}$ - proceeding with a scheme of a lower order.

The value of $\tilde{O}$ is determined on the basis of product $c_p \cdot \overline{O}$ in which the constants $c_p$ are empirically defined and the following values are accepted as suitable: $c_2=.05$, $c_4=.09$, $c_5=.1$, $c_7=.2$.

**Evaluation of the effectiveness.** The advantage of selecting an optimum scheme for numerical integration of ODES, treated from the computing expenses point of view, is different in every separate case. The stability of the RKF schemes is investigated on the basis of the model equations [3] with regulation of the step length.

It is of major importance to know how the use of schemes of a lower precision order influences the global stability of the solution. Numerical experiments have been carried out for integration of six satellite orbits with main semi-axes and eccentricities, given in Table 1. The RKF scheme of 7/8 order is accepted as a guarantee of the numerical solution stability with selection of a constant step within definite limits. The effectiveness is estimated on the basis of the number of computations on the right side of (4) by the formula $Q = \frac{n_7 - n_{\sim}}{n_7}$ where $n_7$ refers to the guaranteed scheme, and $n_{\sim}$ - to the case of choosing an optimum scheme.

Table 1.

| Orbit number | Eccentricity | Main semi-axis [m] |
|---|---|---|
| 1 | .01 | 8 000 000 |
| 2 | .02 | 10 000 000 |
| 3 | .04 | 20 000 000 |
| 4 | .05 | 30 000 000 |
| 5 | .07 | 42 241 085 |
| 6 | .08 | 50 000 000 |

Figure 1 presents the effectiveness of the computations for each separate orbit at different integration steps as well as the total effectiveness. The stability of the numerical solution is estimated on the basis of equation of the type of (4), which includes different disturbing factors, related to the Earth gravitation field. Figure 2 presents the differences, obtained in the selection of optimum integration scheme, towards the use of a scheme of the maximum order. The temporal integration interval is one day and night. The computations are made with requirements for local relative error $\varepsilon \leq 2.10^{-7}$. The estimation of the results allows to draw up the following basic conclusions:

- the choice of a scheme, based on the error estimation with the RKF methods can produce considerable economy of machine time;
- by decreasing the integration step, the global stability of the solution decreases with using schemes of lower precision order, but it is completely satisfactory for a wide range of problems; when necessary to achieve a better stability for a definite class of orbits, it is necessary to point out a smaller local error;
- when multi-parameter computations are made in the field of the immitation modelling, the experimenter can trust the proposed strategy for selection of computation schemes.

The numerical experiments were carried out on the basis of programmes using the Fortran programme language [24]. For this purpose an integrator of systems of independent vector differential equations has been developed. It allows the integration of every system to be performed independently from the integration of the remaining ones, with individual choice of a method of appropriate order.

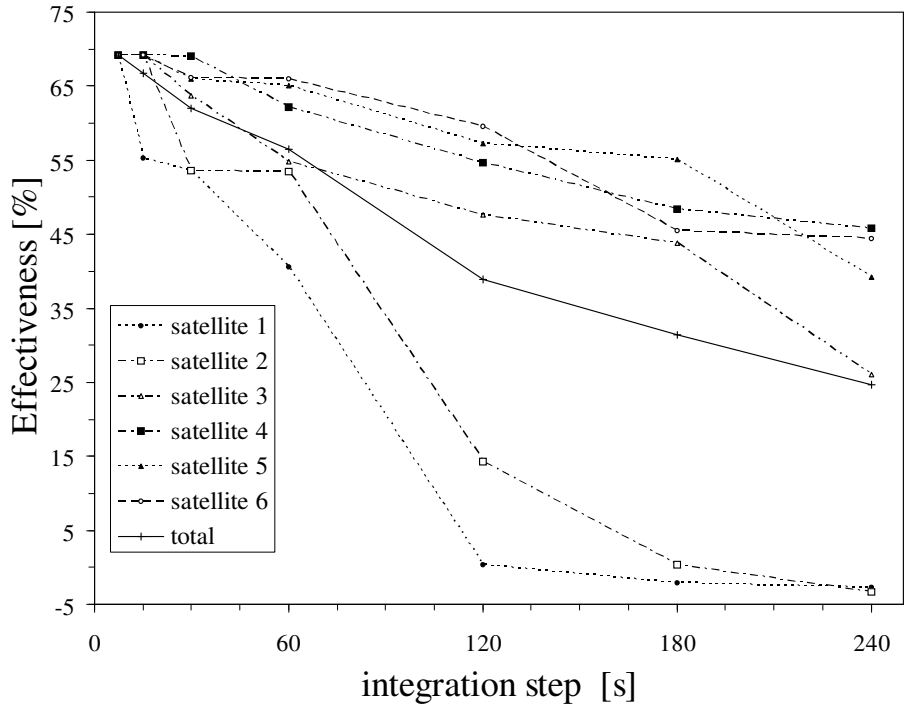

Fig. 1. Average efficiency for one day and night with numerical integration of the equation for the Artificial Earth satellites motion with selection of an optimum scheme.

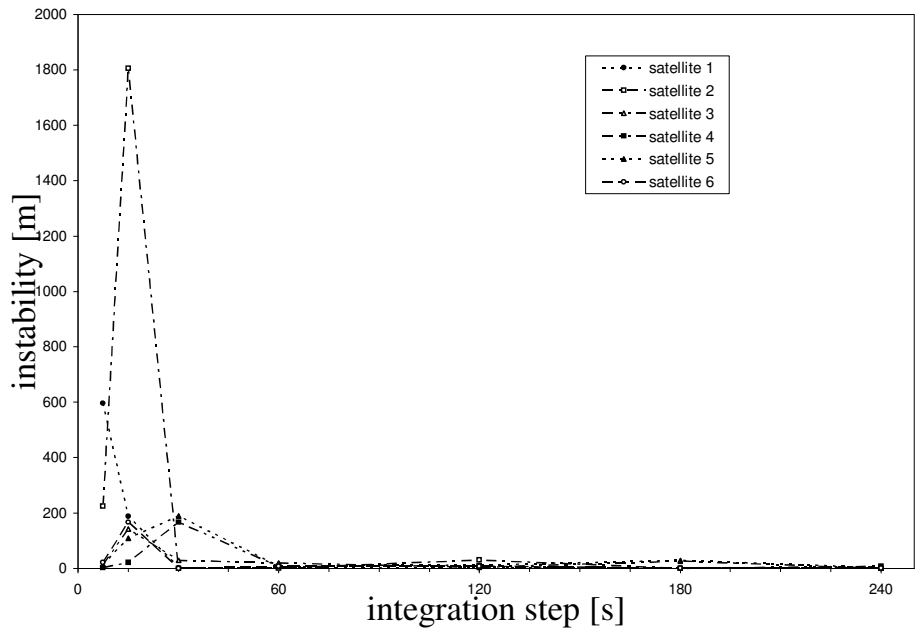

Fig. 2. The difference between the solution, obtained with selection of an optimum scheme and with a scheme of the 7$^{th}$ order, obtained for one day and night at $\varepsilon \leq 2.10^{-7}$

The examined approach for integration of independent differential equations systems is applicable for solution of problems with many objects. The enhancement of the computational efficiency is especially beneficial for problems, whose solution should be made in real time. The examined approach affords additional opportunities, connected with computational parallelization in multi-processor systems as well as in network computations.